\documentclass[pra,twocolumn, showpacs,showkeys,secnumarabic,amssymb,amsmath,superscriptaddress]{revtex4-1}
\usepackage{amssymb,amsmath}
\usepackage{graphicx}
\usepackage{bm}
\usepackage{color}
\begin{document}

\title{Quantum coherence of the Heisenberg spin models with Dzyaloshinsky-Moriya interactions}

\author{Chandrashekar Radhakrishnan }
\email{Corresponding author: chandrashekar.radhakrishnan@nyu.edu}
\affiliation{New York University Shanghai, 1555 Century Ave, Pudong, Shanghai 200122, China}
\affiliation{NYU-ECNU Institute of Physics at NYU Shanghai, 3663 Zhongshan Road North, Shanghai 200062, China}

\author{Manikandan Parthasarathy}
\affiliation{Department of Physics, Ramakrishna Mission Vivekananda College, Mylapore, Chennai 600004, India}

\author{Segar Jambulingam}
\affiliation{New York University Shanghai, 1555 Century Ave, Pudong, Shanghai 200122, China}
\affiliation{Department of Physics, Ramakrishna Mission Vivekananda College, Mylapore, Chennai 600004, India}

\author{Tim Byrnes}
\email{Corresponding author: tim.byrnes@nyu.edu}
\affiliation{State Key Laboratory of Precision Spectroscopy, School of Physical and Material Sciences,
East China Normal University, Shanghai 200062, China}
\affiliation{New York University Shanghai, 1555 Century Ave, Pudong, Shanghai 200122, China}
\affiliation{NYU-ECNU Institute of Physics at NYU Shanghai, 3663 Zhongshan Road North, Shanghai 200062, China}
\affiliation{National Institute of Informatics, 2-1-2 Hitotsubashi, Chiyoda-ku, Tokyo 101-8430, Japan}
\affiliation{Department of Physics, New York University, New York, NY 10003, USA}

\begin{abstract}
We study quantum coherence in a spin chain with both symmetric exchange and antisymmetric Dzyaloshinsky-Moriya couplings. Quantum coherence is quantified using the recently introduced quantum Jensen-Shannon divergence, which has the property that it is easily calculable and has several desirable mathematical properties. We calculate exactly the coherence for arbitrary number of spins at zero temperature in various limiting cases.  The $ \sigma^z \sigma^z $ interaction tunes the amount of coherence in the system, and the antisymmetric coupling changes the nature of the coherence.  We also investigate the effect of non-zero temperature by looking at a two-spin system and find similar behavior, with temperature dampening the coherence.  The characteristic behavior of coherence resembles that of entanglement and is opposite to that of discord. The distribution of the coherence on the spins is investigated and found that it arises entirely due to the correlations between the spins.  
\end{abstract}


\maketitle

%
%
%
%
Quantum coherence is one of the central concepts in quantum physics, and hence its detection and quantification is a fundamental task.  Traditionally, the distinction between quantum and classical coherence is made using 
phase space distributions \cite{glauber1963coherent,sudarshan1963equivalence} and higher order correlation functions
\cite{scully1997quantum}.  While this gives some insight into the nature of coherence, these techniques do not quantify coherence in a rigorous sense.  Recently a scheme for measuring coherence was developed by Baumgratz and co-workers in Ref. \cite{baumgratz2014quantifying} based on the framework of quantum information theory. 
Fundamental quantities such as incoherent states, maximally coherent states and incoherent operations which are needed for the 
development of the procedure were introduced and analyzed.  Rapid developments have been made in understanding the theory of quantum 
coherence through numerous works \cite{yao2015quantum,yadin2015quantum,ma2016converting,zhang2016quantifying,xu2016quantifying,yadin2016general,bromley2015frozen,
du2015conditions,pires2015geometric,cheng2015complementarity,mani2015cohering,killoran2016converting} and 
in using it as a resource in quantum information theory \cite{streltsov2016quantum,Marvian2016how,winter2016operational,brandao2015reversible,chitambar2016critical,chitambar2015relating,
del2015resource}.  Investigations on the role of quantum coherence in thermodynamic processes \cite{lostaglio2015description,narasimhachar2015low,lostaglio2015quantum}, assisted subspace discrimination \cite{zhang2016intrinsic}, quantum state merging \cite{streltsov2016Entanglement} and in the generation of 
gaussian entanglement \cite{wang2016gaussian} have been carried out.  Currently there is a lot of interest in applying the procedure of quantifying coherence
and relating them to experimental quantities in feasible systems like Bose-Einstein condensates \cite{opanchuk2016quantifying,pyrkov2014full}, cavity optomechanical system \cite{zheng2016detecting,man2015cavity} and 
spin systems \cite{karpat2014quantum,malvezzi2016quantum,chen2016coherence,li2016quantum,cheng2016finite,ccakmak2015factorization}.  
 
In Ref.  \cite{baumgratz2014quantifying} the set of properties a functional should satisfy in order to be 
considered as a coherence measure were proposed.  Based on these developments several functions were introduced to serve as measures of
coherence \cite{baumgratz2014quantifying,shao2016quantum,qi2016coherence,yao2016frobenius,napoli2016robustness,rana2016trace,
rastegin2016quantum,adesso2016measures,radhakrishnan2016distribution}.  All these measures of coherence can be broadly classified into either the entropic or the geometric class of measures.
This depends on whether the measure is based on the entropy functional or has a metric nature which can give rise to a 
geometric structure.  A mathematical functional is deemed to be a metric if it obeys the axioms of distance and satisfies the 
triangle inequality.  In Ref. \cite{radhakrishnan2016distribution} we introduced a new measure based on the quantum version of
the Jensen-Shannon divergence.  It was found previously that this measure has many desirable mathematical properties as it combines the features of both the entropic and geometric class. For example, the quantum Jensen-Shannon divergence is a distance measure, is symmetric in its input states, its square root obeys the triangle inequality, and is easily calculable. These properties allowed us to decompose the total coherence in a system into various contributions.  Coherence can originate from individual subsystems (local coherence) or may be
distributed across several subsystems (intrinsic coherence). 
Intrinsic coherence has the property that local basis transformations, which in general change the amount of coherence (coherence is a basis dependent quantity), leave the coherence unaffected.  Several models were examined in Ref. \cite{radhakrishnan2016distribution}, but to date utilizing the quantum Jensen-Shannon divergence for quantifying coherence has not been explored any further.

In the context of quantum many-body systems, it was found in numerous works that entanglement 
can be used to detect quantum phase transitions in condensed matter systems \cite{osterloh2002scaling,osborne2002entanglement,vidal2003entanglement,wu2004quantum}. This is natural since quantum 
correlations underlies both entanglement and quantum phase transitions.  But entanglement accounts only for non-local quantum 
correlations and does not consider other kinds of correlations and 
quantum features \cite{henderson2001classical,ollivier2001quantum}.  To have a complete understanding of the 
role played by quantumness of an object in physical phenomena it is important to consider the role played by other features like 
quantum coherence.  Studies have been carried out with a view to understand the role of quantum coherence in spin chain models.
Most of these works were focussed on anisotropic XY model \cite{karpat2014quantum} with a view to understand the role of 
quantum coherence in investigating quantum phase transitions. 
Such a deep understanding of coherence is applicable to not only naturally occurring systems in condensed matter, but also artificially created metamaterials such as those being routinely produced with cold atom, ion trap, and superconducting systems \cite{salathe2015digital,buluta2009quantum,georgescu2014quantum,buluta2011natural}. Of particular interest are materials with spin-orbit coupling as this is one of the basic ingredients of topological quantum states, which are fundamental to many fascinating effects such as the quantum Hall effect, topological insulators, and anyonic statistics.  

In this paper, we study the behavior of quantum coherence in the XYZ spin chain with Dyzaloshinskii-Moriya (DM) interactions.  The Hamiltonian that we consider is
\begin{align}
H = & \sum_{n} J_{x} \sigma^{x}_{n} \sigma^{x}_{n+1} + J_{y}   \sigma^{y}_{n} \sigma^{y}_{n+1} 
      + J_{z} \sigma^{z}_{n} \sigma^{z}_{n+1} \nonumber \\
		& 	+  \vec{D}.(\vec{\sigma}_{n} \times \vec{\sigma}_{n+1})
\label{hdmhamiltonian}       
\end{align}
where $ J_{x,y,z} $ are the symmetric exchange spin-spin interactions, $\vec{D} $ is the antisymmetric DM exchange interaction, and 
$ \sigma^{x,y,z}_i $ are Pauli spin operators on the site $ i $. When $J_{\ell} > 0$, ($\ell = x,y,z$) the system is antiferromagnetic in nature and for $J_{\ell} < 0$ it has ferromagnetic
behaviour.  The model described above is integrable and an exact solution can be obtained using Bethe Ansatz
\cite{kirillov1987exact,inami1994integrable}.  Despite its simplicity
the model describes the magnetic properties of a wide range of compounds and also has a very rich mathematical structure based on 
group theory \cite{bonechi1992heisenberg,baker1964high}. Though the Heisenberg model was successful in describing the magnetism 
of several systems like $\text{KCuF}_{3}$, $\text{CsNiCl}_{3}$ and $\text{CsCuCl}_{3}$ it could not explain the 
weak ferromagnetic behaviour of certain compounds like $\alpha$-$\text{Fe}_{2} \text{O}_{3}$, 
$\text{MnCO}_{3}$ and $ \text{CoCO}_{3}$ in their antiferromagnetic state.  A detailed 
investigation of this problem was carried out by Dyzaloshinskii \cite{dzyaloshinsky1958thermodynamic} and Moriya
\cite{moriya1960anisotropic}. As described in \cite{dzyaloshinsky1958thermodynamic} weak ferromagnetic behavior arises due to the
relativistic spin-lattice and dipole interactions.  This DM  interaction causes spin canting, a behavior in which the spins
are tilted by a small angle instead of being exactly parallel to each other.  This characteristic feature is the source of the small 
amount of magnetism in the above mentioned materials.  Based on crystal symmetry \cite{moriya1960anisotropic} it was found that spin-canting can be described by the DM antisymmetric exchange interaction term.

Generally, the magnitude of the DM interaction is small compared to the spin-spin interaction in naturally occurring materials, but is important to include it as it changes the nature of the system greatly. These changes have been studied from the perspective of quantum information theory particularly in the measurement
of quantum correlations like entanglement \cite{kargarian2009dzyaloshinskii,li2008thermal,li2008entanglement}
and quantum discord \cite{liu2011quantum,yi2010thermal}.  
From the studies on quantum entanglement \cite{li2008entanglement} and discord \cite{yi2010thermal}, it was found that the 
critical temperature below which these correlations start decreasing can be altered by varying the DM interaction.  Further,
the direction of the DM interaction has a direct bearing on the extent to which this change happens. In particular, it was found 
that as the DM interaction increases, the entanglement increases whereas the quantum discord decreases.

This paper is organized as follows. First we examine the coherence of certain limiting cases of the model which allow 
for an exact solution.  Later  we characterize the
quantum coherence of a two qubit spin models with added DM coupling interaction.  Finally the basis dependence of 
quantum coherence and its effect in spin chain models is presented.  A summary of our conclusions is presented in the 
Section of Discussion.

%
%
%
\section*{Results}
\label{results}
\subsection*{Limiting cases at zero temperature}
\label{Lc}
In this section we consider some of the limiting cases of the Hamiltonian (\ref{hdmhamiltonian}) where it is possible to obtain an exact expression for the coherence at zero temperature.

\subsubsection*{$ J_z = 0 $ case}

The first limiting case we consider is where $ J_z = 0 $, $ J_x = J_y = J $, and $ \vec{D} = (0,0,D) $, which is written
\begin{align}
H = \sum_{n=1}^{N} & J \left(   \sigma_{n}^{x} \sigma_{n+1}^{x} + \sigma_{n}^{y} \sigma_{n+1}^{y} \right)  \nonumber \\
&  + D \left( \sigma_{n}^{x} \sigma_{n+1}^{y} - \sigma_{n}^{y} \sigma_{n+1}^{x} \right).
\label{xydmhamiltonian}  
\end{align}
Being a one-dimensional system, this can be explicitly solved using a Jordan-Wigner transformation to 
map between spins and fermions: 
\begin{eqnarray}
c_{k} &=&  ( \prod_{n=1}^{k-1} 2 \sigma_{n}^{+} \sigma_{n}^{-} - 1) \sigma_{k}^{+}   ,
\label{fermionicmap}
\end{eqnarray}
where the operators defined in (\ref{fermionicmap}) obey the commutation relations for fermions
\begin{equation}
\{c_{k}^{\dagger},c_{l}\} =  \delta_{kl}, \;\; \{c_{k},c_{l}\}  = 0.
\label{anticommutation}
\end{equation}
Under this transformation, the Hamiltonian (\ref{xydmhamiltonian}) can be recast
as
\begin{align}
H =  \sum_{i=1}^N \frac{J}{2}(c_{i+1}^{\dagger} c_{i} + c_{i}^{\dagger} c_{i+1}) + 
    \frac{D}{2i}(c_{i}^{\dagger} c_{i+1} - c_{i+1}^{\dagger} c_{i}) .
\label{fermionicbasisHamiltonian}    
\end{align}
For periodic boundary conditions this can be diagonalized by Fourier transform where $ c_{q} = \frac{1}{\sqrt{N}}\sum_n e^{inq} c_n $. 
\begin{equation}
H = \sum_{q} \Lambda_{q} c_{q}^{\dagger} c_{q},
\label{momentumspaceHamiltonian}
\end{equation}
where 
\begin{align}
\Lambda_{q} & = J \cos(q) + D \sin(q) \nonumber \\
& =  \sqrt{J^2+D^2} \cos (q- \theta) 
\label{energyspectrum} 
\end{align}
and
\begin{align}
\cos \theta = \frac{J}{\sqrt{J^2+D^2}} .
\end{align}
The ground state of the Hamiltonian (\ref{xydmhamiltonian}) is in the total spin $ \sum_n \sigma^z_n = 0 $ sector, which corresponds to having $ N/2 $ fermions.  

The ground state can then be written as
\begin{align}
| \Psi_0 (J_z=0) \rangle = & \prod_{\Lambda_{q} < 0  } c^{\dagger}_{q} |0 \rangle \nonumber \\
 = &\frac{1}{N^{N/4}} \sum_{n_1} \sum_{n_2} \dots \sum_{n_{N/2}} e^{-i \bm{n} \cdot \bm{q}} 
\nonumber \\
& \times c_{n_1}^\dagger 
 c_{n_2}^\dagger \dots  c_{n_{N/2}}^\dagger | 0 \rangle
\label{arbitrarystate}
\end{align}
where $ \bm{n} = (n_1, n_2, \dots , n_{N/2}) $ and $ \bm{q} = (q_1, q_2, \dots , q_{N/2}) $ are the sets of momentum satisfying 
$ \Lambda_{q} < 0  $.  Reverting to the spin formulation, the ground state is
\begin{align}
| \Psi_0  (J_z=0)\rangle = & \frac{1}{\sqrt{{N \choose N/2}}} \sum_{\sigma_1, \dots , \sigma_N; \sum_n \sigma_n = 0 } e^{-i \Omega( \sigma_1, \dots , \sigma_N)} \nonumber \\
& \times| \sigma_1, \dots , \sigma_N \rangle 
\label{groundstate}
\end{align}
where $ \Omega( \sigma_1, \dots , \sigma_N) $ is a phase factor which originates from (\ref{arbitrarystate}) and the phases picked up by the Jordan-Wigner transformation.  

We now evaluate the coherence using the square root of the quantum Jensen-Shannon divergence, defined as 
\begin{align}
C  & =
\sqrt{\mathcal{J}(\rho,\rho_{d})}  \nonumber \\
                            &  = \sqrt{S \left( \frac{\rho + \rho_{d}}{2}\right) 
                                  - \frac{S(\rho)}{2}  -  \frac{S(\rho_{d})}{2}} . 
\label{qjsdistance} 
\end{align} 
This was introduced in Ref. \cite{radhakrishnan2016distribution} as a measure of coherence.  Here $\rho$ and $\rho_{d}$ denote the density matrix of the state and the closest incoherent state under the distance measure and $S(\rho)$ is the 
von Neumann entropy. We take the closest incoherent state to be the density matrix where all the off-diagonal elements of $\rho$ 
are set to zero  \cite{baumgratz2014quantifying}. 
For pure states this measure satisfies all the axioms of a distance and also obeys the triangle
inequality and hence is a metric. Thus this measure has both entropic nature and geometric properties.  In the limit where $N$ is sufficiently large, 
the quantum coherence of the system as measured using the Jensen-Shannon divergence (\ref{qjsdistance}) attains the value $C=1$ as shown below.  To evaluate the measure 
we need to find the entropy of the density matrix of the state, its closest incoherent state $\rho_{d}$ and also the entropy of the symmetric sum of these density matrices.  

  Firstly, since the state is pure, the entropy of the density matrix $\rho$ is zero. For the state (\ref{groundstate}) the precise knowledge of  $ \Omega( \sigma_1, \dots , \sigma_N) $ is in fact not necessary, as it is invariant under the particular phases of the state.  
  We may thus take the phases to be zero, and we evaluate the eigenvalues of the incoherent density matrix and the $(\rho+\rho_{d})/2$ to be 
\begin{eqnarray}
\lambda^{i}(\rho_{d}) &=& \frac{1}{{N \choose {N/2}}},  \forall i \in \{1,{N \choose {N/2}}\} \\
\lambda^{i}\left(\frac{\rho+\rho_{d}}{2} \right) &=&  
           \begin{cases} 
              \frac{{N \choose {N/2}}+1}{2 {N \choose {N/2}}}, & \text{if} \ i=1 \\
              \frac{1}{2 {N \choose {N/2}}}, & \text{if} \ i \in \{2, {N \choose {N/2}}\}.
           \end{cases}
\label{evalsdensitymatrix}                                       
\end{eqnarray}
From the eigenvalues we find the entropy of the $(\rho+\rho_{d})/2$ to be
\begin{eqnarray}
S\left(\frac{\rho+\rho_{d}}{2}\right) &=& - \frac{{N \choose {N/2}} +1}{2 {N \choose {N/2}}} 
                                \log_{2} \left(\frac{{N \choose {N/2}} +1}{2 {N \choose {N/2}}}\right) \nonumber \\
                                      & &  - \frac{{N \choose {N/2}}-1}{2 {N \choose {N/2}}} 
                                      \log_{2} \left(\frac{1}{2 {N \choose {N/2}}}\right)
\label{entropyrhorhod}                                      
\end{eqnarray}
and the entropy of the incoherent state is $S(\rho_{d}) = \log_{2}{N \choose {N/2}}$. Our final expression for the coherence 
is then
\begin{align}
C ( J_{z}=0) &=  \sqrt{ 1 + \frac{ \log_2 m}{2} - \frac{1}{2}( 1 + \frac{1}{m}) \log_2 (m+1) }																			\label{jzzerolimit}
\end{align}
%
%
where $ m = {N \choose {N/2}} $. The quantum coherence approaches 1 for large $N$, which is the maximum value of the the Jensen-Shannon divergence. 
The large amount of coherence of this state originates from the superposition of ${N \choose {N/2}} $ terms in  (\ref{groundstate}).  The phase factor $\Omega$ in Eqn. (\ref{groundstate}) originating from the Fourier transform and the 
Jordan-Wigner transformation changes the nature (i.e. the basis) of the coherence.  These phases depend on both the spin-spin 
interaction parameter and the DM coupling.

\subsubsection*{$ J_{x,y} = D = 0 $ case}

The opposite limit is in the ferromagnetic limit with $ J_{x,y} = D = 0 $, with Hamiltonian
\begin{equation}
H = J \sum \sigma_{i}^{z} \sigma_{i+1}^{z} .
\label{ferromagneticlimit}
\end{equation}
For $J < 0$ for ferromagnetism, the ground state configuration of the system is either
\begin{align}
\Psi_0 (J_{x,y} = D = 0) \rangle = | \sigma, \sigma, \dots , \sigma \rangle 
\label{ferro}
\end{align}
with $ \sigma = \pm 1 $.  For $J > 0$ we have antiferromagnetism, and the ground state is a Neel state with every second spin flipped in (\ref{ferro}).  In either case, we take the condensed matter physics point of view that the ground state is spontaneously broken to one of these states. In all these cases the states are already in the decohered basis, hence
\begin{align}
C ( J_{x,y} = D = 0 ) = 0   .  
\label{jzlargelimit}
\end{align}
If we choose to preserve the spin-flip symmetry the ferromagnetic ground state is 
\begin{align}
\Psi_0 (J_{x,y} & = D = 0) \rangle = \nonumber \\
& \frac{1}{\sqrt{2}} (| 1, 1, \dots , 1 \rangle + | -1, -1, \dots , -1 \rangle ) .
\label{antiferro}
\end{align}
here the coherence evaluates to
\begin{align}
C  ( J_{x,y} = D = 0 ) = 0.56  .  
\label{jzlargelimit2}
\end{align}
In either case the coherence is considerably lower than the opposite regime as given in (\ref{jzzerolimit}). 
We thus see the general behavior that for zero $J_z$ the zero temperature ground state has a large amount of coherence, as given by (\ref{jzzerolimit}).  As the $J_z$ term is increased, this diminishes, and for the spontaneously broken case of (\ref{jzlargelimit}) it reduces to zero.  Thus the $J_z$ parameter can be seen as the crucial control for the amount of coherence.

%
%
%
\subsection*{Two-site model at non-zero temperature}
\label{xyz}

The previous section revealed the general behavior that is expected in the XYZ model with DM interactions.  We now examine what occurs at intermediate values of the parameters, and also include non-zero temperature effects. An exact solution for the thermodynamic limit is not available for the general case.  We therefore consider a two-site system which should show the general dependence of the coherence on various parameters. The Hamiltonian of a two qubit Heisenberg model with Dzyaloshinsky Moriya interaction is 
\begin{equation}
H = J_{x} \sigma_{1}^{x} \sigma_{2}^{x} + J_{y} \sigma_{1}^{y} \sigma_{2}^{y} + J_{z} \sigma_{1}^{z} \sigma_{2}^{z} 
    + {\vec{D}}.(\vec{\sigma_{1}} \times \vec{\sigma_{2}}).
\label{xyzdmHamiltonian}    
\end{equation}
The DM interaction can be oriented in an arbitrary direction in general. We examine the effect of the DM interaction in various directions in the following sections.

\subsubsection*{DM interaction in the z-direction}
\label{xyzdz}

The Hamiltonian of a two qubit anisotropic $XYZ$ model with DM interaction oriented along the $z$-direction $\vec{D} = (0,0,D_z) $ is 
\begin{equation}
H = J_{x} \sigma_{1}^{x} \sigma_{2}^{x} + J_{y} \sigma_{1}^{y} \sigma_{2}^{y} + J_{z} \sigma_{1}^{z} \sigma_{2}^{z} + D_{z}(\sigma_{1}^{x} \sigma_{2}^{y} - \sigma_{1}^{y} \sigma_{2}^{x}),
\label{xyzdzhamiltonain}
\end{equation}
The model is ferromagnetic when $J_{x,y,z}< 0 $ and antiferromagnetic when $J_{x,y,z}> 0 $. 
Diagonalizing the Hamiltonian in (\ref{xyzdzhamiltonain})  the eigenvalues are
\begin{align}
E_{1,2} & = J_{z} \pm J_{x} \mp J_{y}, \nonumber \\
E_{3,4} & = - J_{z}  \pm \omega.
\end{align}
The eigenvectors of the Hamiltonian are
\begin{align}
|\Psi_{1,2} \rangle & = \frac{| \downarrow \downarrow  \rangle \pm |\uparrow \uparrow \rangle}{\sqrt{2}} \nonumber \\
|\Psi_{3,4} \rangle & = \frac{|\downarrow  \uparrow \rangle \pm e^{-i\theta} |\uparrow \downarrow  \rangle}{\sqrt{2}},
\label{evectorxyzdz}
\end{align}
where 
\begin{eqnarray*}
\cos \theta  &=& \frac{(J_{x}+J_{y})/2}{\sqrt{D_{z}^{2} + (\frac{J_{x} + J_{y}}{2})^{2}}} \\
\omega &=& \sqrt{4D_{z}^{2} + (J_{x} + J_{y})^{2}}.  
\end{eqnarray*}

Here we are interested in investigating the temperature dependence of the quantum coherence in the two qubit 
Heisenberg model.  This can be accomplished from the knowledge of the thermal density matrix $\rho(T)$ of the model.  The thermal
density matrix $\rho(T) = \exp(- \beta H)/Z$ describes the state of a spin chain at thermal equilibrium, where 
$Z = \text{Tr} [\exp (-\beta H)]$ is the partition function of the system and $\beta = 1/k_{B} T$ with $k_{B}$ and $T$ being the Boltzmann
constant and the temperature respectively.  Throughout our discussion we assume $k_{B} =1$ for the sake of convenience.  In the 
computational basis (i.e. $ \sigma^z $ eigenstates) the thermal density matrix of the two qubit $XYZ$ model with DM interaction is
\begin{equation}
\rho(T) = \frac{1}{Z}\left(
\begin{array}{cccc}
r&0&0&s\\
0&u&v&0\\
0&v^{*}&u&0\\
s&0&0&r\\
\end{array}
\right)
\label{thermaldmxyzdz}
\end{equation}
where the matrix elements are
\begin{align}
r& =e^{- J_{z}/T} \cosh \left(\frac{J_{x}-J_{y}}{T}\right) \nonumber \\
u& =e^{J_{z}/T} \cosh \left(\frac{\omega}{T} \right)  \nonumber \\
v& = - \frac{e^{J_{z}/T}}{\omega} \sinh \left(\omega/T\right)(2i D_{z}+J_{x}+J_{y})  \nonumber \\
s& =e^{- J_{z}/T} \sinh \left(\frac{J_{x}-J_{y}}{T} \right).
\end{align}
The partition function of the system is then
\begin{align}
Z = 2 e^{-J_{z}/T}  \cosh \left(\frac{J_x - J_{y}}{T}\right) + 2 e^{ J_{z}/T}  \cosh (\frac{\omega}{T}).
\end{align}

\begin{figure}[t]
\begin{center}
\includegraphics[width=\columnwidth]{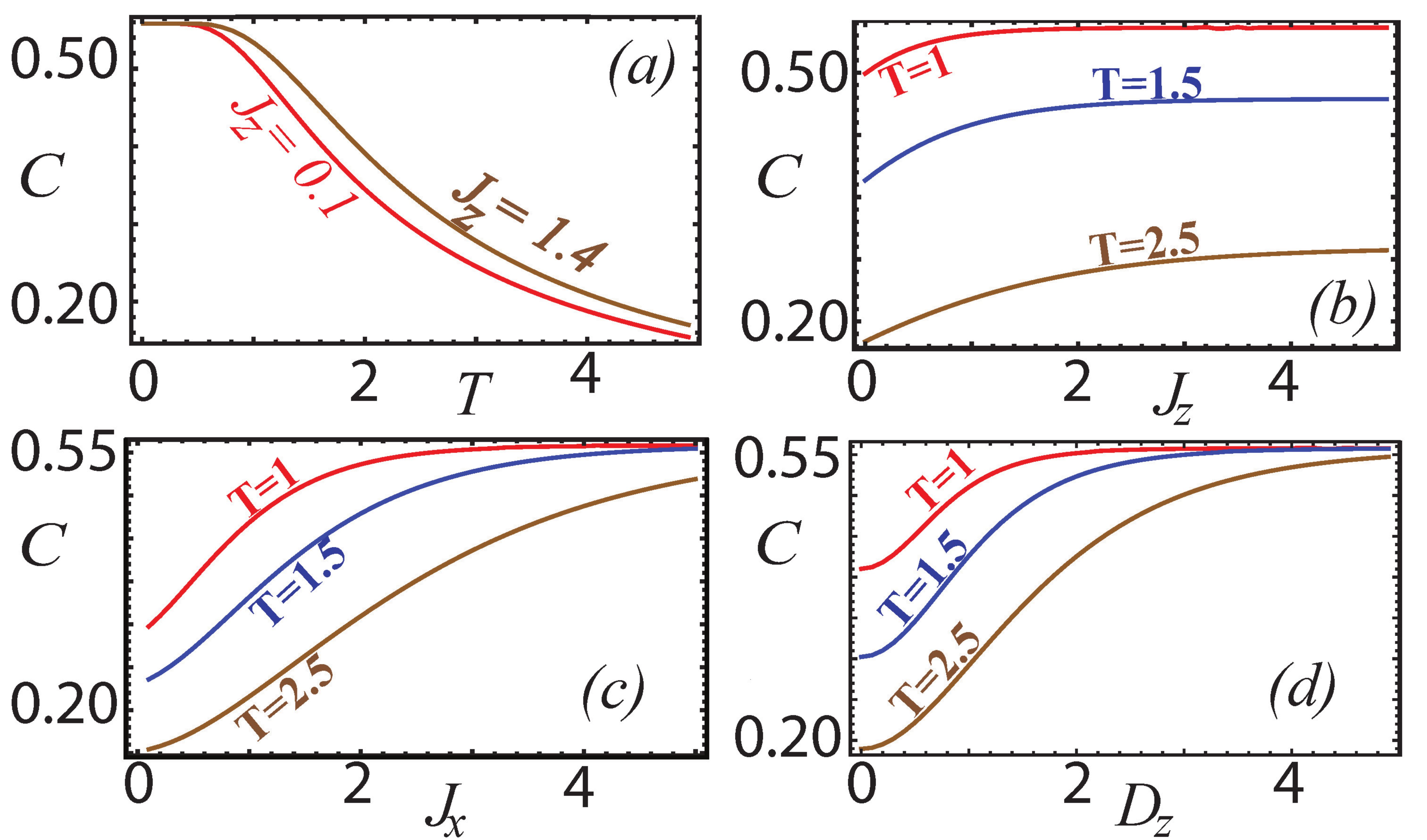}
\end{center}
\caption{(a) Quantum coherence versus temperature for different $J_{z}$, for $\vec{D} = (0,0,1) $, $J_x =-1$ and $J_y = -0.5$. (b) 
Quantum coherence versus $J_{z}$ for different temperatures for $\vec{D} = (0,0,1) $, $J_x  =  -1$ and $J_y = -0.5$. (c) Quantum 
coherence versus $J_{x}$ for different temperatures for $\vec{D} = (0,0,1) $, $J_{z} = -1$ and $J_{y} = -0.5$. (d) Quantum coherence 
versus $D_z $ for different temperatures when $J_{z} = 0.2$ $J_{x} = -1$ and $J_{y} = -0.5$.}
\label{fig1}
\end{figure}

Fig. \ref{fig1} gives the typical behavior of quantum coherence for the $XYZ$ model including the DM interaction in the $ z $ direction. 
We first observe that coherence decreases with temperature as expected due to an increase in 
thermal fluctuations, which has the effect of decreasing quantum coherence (Fig. \ref{fig1}(a)).
All the quantum correlations are generally constant at very low temperature and starts to decay only after a 
thershold temperature.  This is because thermal fluctuations are strong enough to affect the
quantum correlations in the system only above the characteristic temperature set by the gap energy, which is non-zero for a finite sized system. 
We find that the temperature at which the coherence falls from its 
maximal value, tends to increase with interaction parameters.  Previously it has been observed that entanglement displays
a behavior which is similar to the one described above for coherence \cite{li2008thermal,li2008entanglement}.  On the other hand, quantum discord was found to decrease with the interaction parameters for a given temperature \cite{yi2010thermal}.  This difference in behavior between entanglement and discord has been 
argued to be due to the faster decrease of local quantum correlations which are not accounted for by entanglement.  In this 
context we wish to note that due to spin-flip symmetry, the local coherence as defined in \cite{radhakrishnan2016distribution} 
is zero and the total coherence observed in the 
system is entirely due to the correlations between the spins.  Due to the non-local nature of quantum coherence its behavior is 
reminiscent of entanglement in the system.  This gives credence to the point of view that coherence which is not localized 
on any of the qubits may contribute to the entanglement between 
them \cite{radhakrishnan2016distribution,streltsov2015measuring}. 
As the interaction parameters are increased the coherence saturates 
to a fixed value.  This fixed value varies with temperature when $J_{z}$ is varied, but 
when $J_{x}$ and $D_{z}$ are changed the coherence always saturates to the value $C=0.56$. This is because in the 
$z$-basis the $\sigma_{z}$ term contributes to the diagonal elements, whereas $\sigma_{x}$ and $\sigma_{y}$ contributes to 
the off-diagonal terms and hence actively participate in generating coherence. 

\begin{figure}[t]
\begin{center}
\includegraphics[width=\columnwidth]{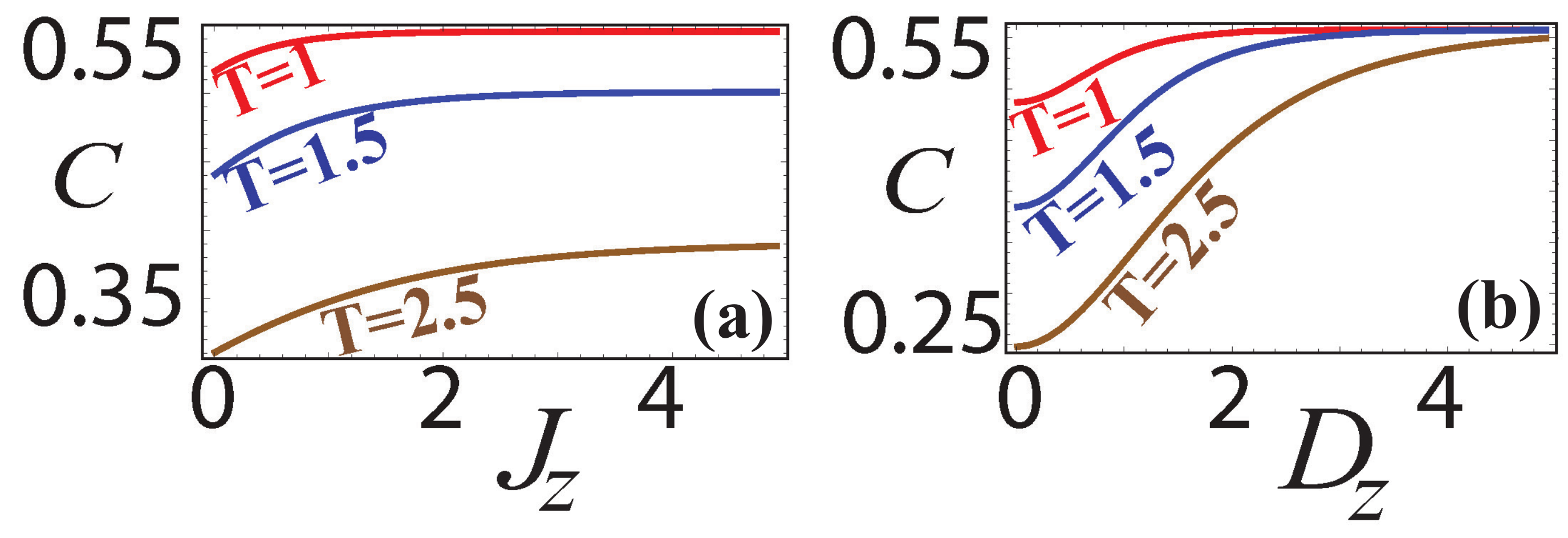}
\end{center}
\caption{Quantum coherence with various parameters in the two site XXZ model with DM interactions in the $ z $-direction. 
(a) Dependence on $J_{z}$ for different temperature with $D_{z}= 1$ and $J_{x}=J_{y}=-1$.  (b) Dependence on $D_{z}$ for 
different temperature with $J_{z} = 1$ and $J_{x} = J_{y} = -1$. }
\label{fig2b}
\end{figure}

 This behaviour is also observed in 
the case of an XXZ model, where the spin-spin exchange coefficients in two directions 
coincide whereas in the third direction it is different ($J_{x} = J_{y} \neq J_{z}$) and the 
DM interaction is in the $z$-direction. Here again we find that the coherence saturates at a finite value.  This 
value varies with temperature when $J_{z}$ is varied, but when $D_{z}$ is changed the saturation value is always 
$C=0.56$.  These conclusions can be observed from the fact that the
 results displayed in plots (a) and (b) of Fig. \ref{fig2b} 
 are identical to the plots (b) and (d) respectively of Fig. \ref{fig1}. Thus we find that the qualitative behaviour
of coherence as a function of the spin-spin and DM interaction parameters shows some universal features in the spin models.

\subsubsection*{DM interaction in the $x$ and $y$-directions}
\label{xyzdy}

We now turn to the two qubit Heisenberg model with DM coupling in the $ x $ and $y$-directions $\vec{D} = (D_x,0,0) $ and $(0,D_y,0) $ respectively. This has the Hamiltonians
\begin{equation}
H = J_{x} \sigma_{1}^{x} \sigma_{2}^{x} + J_{y} \sigma_{1}^{y} \sigma_{2}^{y} + J_{z} \sigma_{1}^{z} \sigma_{2}^{z} + D_{x}(\sigma_{1}
^{y} \sigma_{2}^{z} - \sigma_{1}^{z} \sigma_{2}^{y})
\label{xyzdxhamiltonain}
\end{equation}
for the $ x $-direction and
\begin{equation}
H = J_{x} \sigma_{1}^{x} \sigma_{2}^{x} + J_{y} \sigma_{1}^{y} \sigma_{2}^{y} + J_{z} \sigma_{1}^{z} \sigma_{2}^{z} 
+ D_{y}(\sigma_{1}^{z} \sigma_{2}^{x} - \sigma_{1}^{x} \sigma_{2}^{z}) .
\label{xyzdyhamiltonain}
\end{equation}
for the $ y $-direction.  These have rather similar properties hence we will discuss (\ref{xyzdyhamiltonain}) primarily and simply show the results for (\ref{xyzdxhamiltonain}).  

The eigenvalues of (\ref{xyzdyhamiltonain}) are
\begin{align}
E_{1,2} & = J_{y} \pm J_{x} \mp J_{z}\nonumber, \\
E_{3,4} & = -J_{y} \pm \omega.  
\end{align}
and its eigenstates are
\begin{align}
|\Psi_{1} \rangle &= \frac{(|\downarrow \uparrow \rangle + |\uparrow \downarrow \rangle)}{\sqrt{2}},  \nonumber \\
|\Psi_{2} \rangle &  =  \frac{(|\downarrow \downarrow \rangle - |\uparrow \uparrow \rangle)}{\sqrt{2}}, \nonumber \\
|\Psi_{3} \rangle &=  \sin\phi_{1} \frac{| \downarrow \downarrow \rangle + |\uparrow \uparrow \rangle}{\sqrt{2}}  
                                         - \cos \phi_{1} \frac{|\downarrow \uparrow \rangle - |\uparrow \downarrow \rangle}{\sqrt{2}}, \nonumber \\
|\Psi_{4} \rangle &=  \sin\phi_{2} \frac{|\downarrow \downarrow\rangle + |\uparrow \uparrow \rangle}{\sqrt{2}}  
                           - \cos \phi_{2} \frac{|\downarrow \uparrow \rangle - |\uparrow \downarrow \rangle}{\sqrt{2}}.                                                                                       
\label{evectorxyzdy}                                         
\end{align}
where
\begin{align}
\tan \phi_{1,2} & = \frac{2 D_{y}}{J_{x} + J_{z} \mp \omega}, \nonumber \\
\omega & = \sqrt{4 D_{y}^{2} + (J_{x} + J_{y})^{2}}. 
\end{align}
The thermal density matrix
$\rho(T)$ of the system can then be calculated to be
\begin{equation}
\rho(T) = \left(
\begin{array}{cccc}
m_{1}&-q&q&m_{2} \\
-q&n_{1}&n_{2}&-q \\
q&n_{2}&n_{1}&q \\
m_{2}&-q&q&m_{1} \\
\end{array}
\right)
\label{thermaldmxyzdy}
\end{equation}
where
\begin{align}
m_{1,2} & = \frac{1}{2Z}\left(\pm e^{-\frac{E_2}{T}} +e^{-\frac{E_3}{T}} \sin^{2} \phi_{1} + e^{-\frac{E_4}{T}} \sin^{2} \phi_{2} \right) \nonumber \\
n_{1,2} & = \frac{1}{2Z}\left( e^{-\frac{E_1}{T}} \pm e^{-\frac{E_3}{T}} \cos^{2} \phi_{1}  \pm e^{-\frac{E_4}{T}} \cos^{2} \phi_{2} \right) \nonumber \\
q & = \frac{1}{2Z}\big(e^{-\frac{E_3}{T}} \sin\phi_{1} \cos\phi_{1} + e^{-\frac{E_4}{T}} \sin\phi_{2} \cos\phi_{2} \big).
\end{align}
The partition function of the 
system is  
\begin{align}
Z = 2e^{\frac{-J_{y}}{T}} \cosh \left(\frac{J_x - J_{z}}{T}\right) + 2e^{\frac{J_{y}}{T}} \cosh 
\left(\frac {\omega}{T}\right) .
\end{align}

\begin{figure}[t]
\begin{center}
\includegraphics[width=\columnwidth]{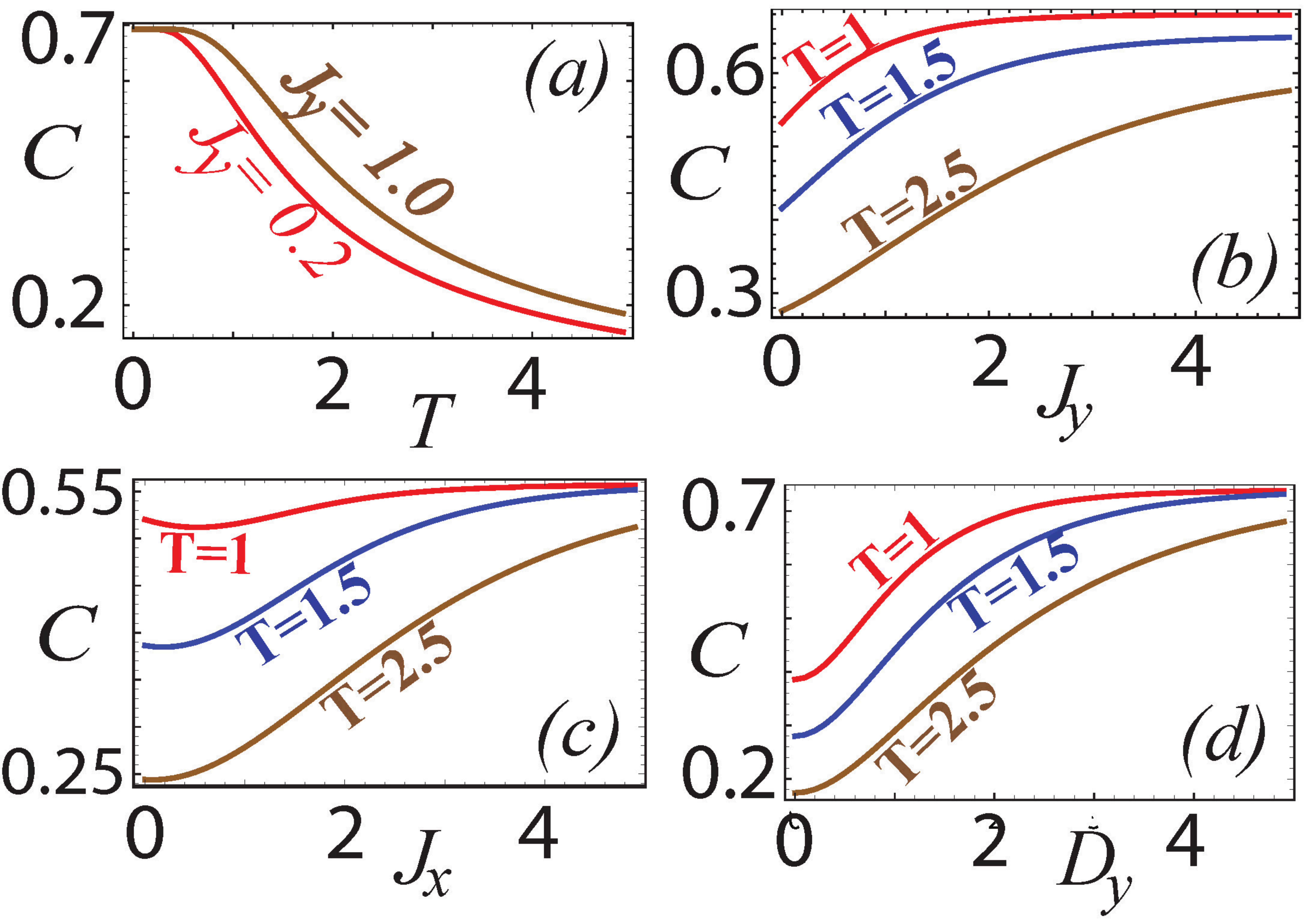}
\end{center}
\caption{Quantum coherence with various parameters in the two site XYZ model with DM interactions in the $ y $-direction. 
(a) Dependence on temperature for different $J_{y}$ with  $D_{y}=1$, $J_{x} = -1$ and $J_{z} = -0.5$. (b) Dependence on $J_y$ for different temperatures with  $D_y = 1$, $J_{x} =-1$ and $J_z  - 0.5$. 
(c) Dependence on $J_x$ for different temperatures with $J_y$ = 0.2 and $J_z$ =-0.5 and 
$D_y = 1$. (d) Dependence on $D_y$ for different temperatures with $J_x = -1$, $J_y = 0.2$ and $J_{z} = -0.5$. The same results are obtained for quantum coherence in the $ x $-direction for parameters $ J_x \rightarrow J_y $, $ J_y \rightarrow J_x $, $ J_z \rightarrow J_z $ and the same temperature.  }
\label{fig3}
\end{figure}

We calculate the coherence in the $ y $-direction using (\ref{qjsdistance}) as before and is displayed in Fig. \ref{fig3}.  
The general trend in the graph indicates that coherence decreases with  
temperature due to the increase in thermal fluctuations which destroys the quantum coherence in the system. 
Coherence increases as the strength of the interaction parameters is increased and saturates to a finite 
value.  From the results we observe that the maximal saturation value is attained when the DM interaction
parameter is varied. 

For coherence in the $ x $-direction we verify that identical results are obtained as that shown in Fig. \ref{fig3}, up to a transformation $ J_x \rightarrow J_y $, $ J_y \rightarrow J_x $, $ J_z \rightarrow J_z $ for the same temperature. This is as expected as in terms of coherence the Hamiltonians (\ref{xyzdxhamiltonain}) and (\ref{xyzdyhamiltonain}) only differ according to a replacement of $ \sigma^x \leftrightarrow \sigma^y $ in the DM interaction.  Both are coherence generating terms and since the coherence is generally unaffected by phase information, similar effects should result from either.

\begin{figure}
\begin{center}
\includegraphics[width=\columnwidth]{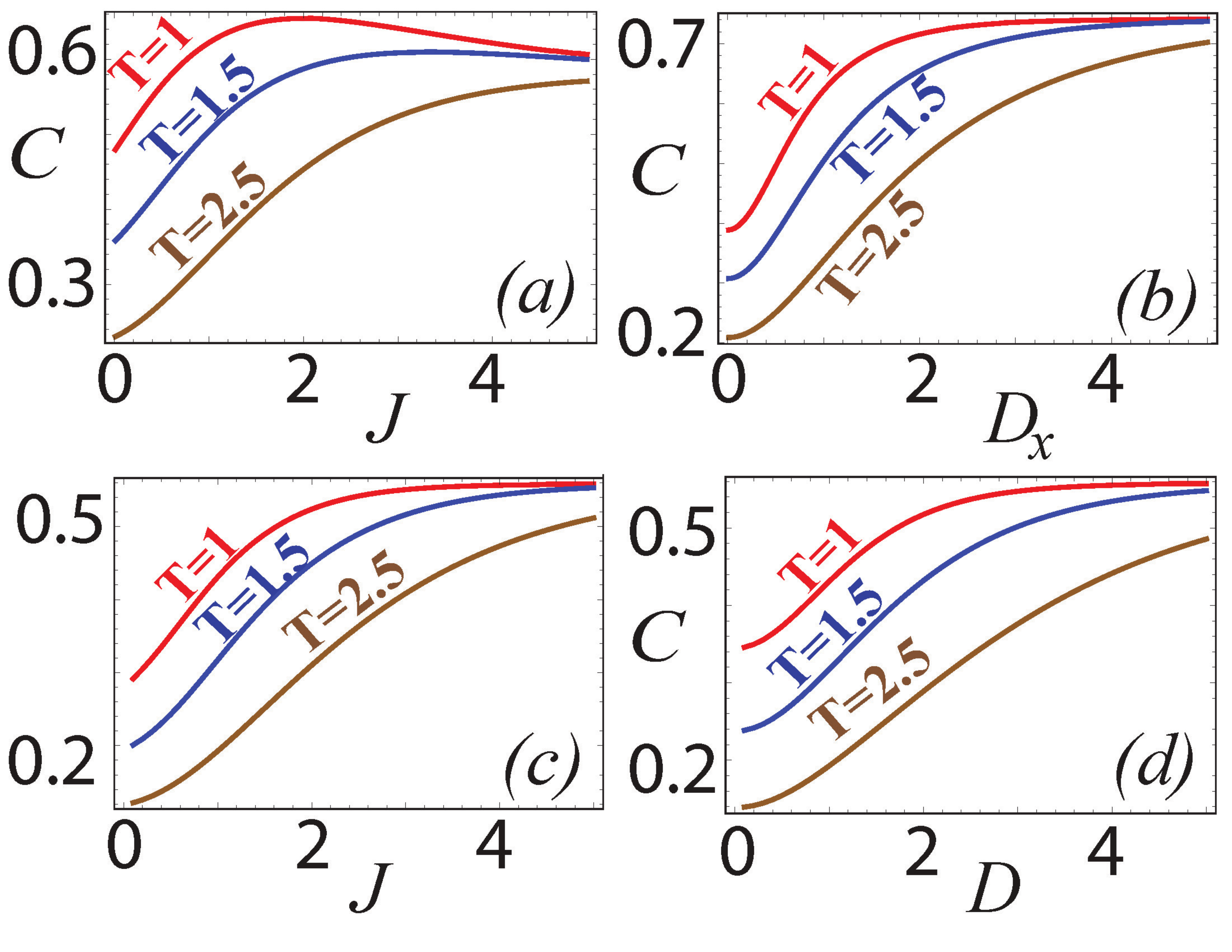}
\end{center}
\caption{Quantum coherence of XXZ model with DM interaction for different temperatures in   
(a) as a function of $J$ for $\vec{D} = (1,0,0) $ and $J_{x} = -1$ and (b) as a function of $D_x$ for 
$J=1$ and $J_{x} = -1$.  Quantum coherence of XXX model with DM interaction for different temperatures
in (c) as a function of $J$ for $\vec{D} = (1,0,0) $ and (d) as a function of $D$ for $J_{x}=J_{y}=J_{z}=-1$. }
\label{fig5}
\end{figure}

The quantum coherence of the XXZ model $(J_{x} = J_{y} \neq J_{z})$ and the 
completely isotropic XXX model $(J_{x} = J_{y} = J_{z})$ with DM interaction 
in the $x$-direction are also analyzed. The behaviour of quantum coherence with respect to 
spin-spin and DM interaction parameters are given in Fig. \ref{fig5} for various temperatures. We observe that the
behaviour of coherence in these models is quite similar to the corresponding results for the XYZ model with 
$D_{y}$ given in Figs \ref{fig3}.  This is because the $ \sigma^x$ term is 
a coherence generating term and has similar effects on both the XYZ Hamiltonian as well as the XXZ model and 
isotropic XXX model.

%
%
%
\subsection*{Basis dependence of quantum coherence}
\label{bsdqc}
Measurement of coherence in quantum systems via measures like relative entropy, $\ell_{1}$-norm 
\cite{baumgratz2014quantifying} and Jensen-Shannon
divergence \cite{radhakrishnan2016distribution} requires a characterization of incoherent states (states with zero coherence).  
To define the incoherent states 
in a $d$-dimensional Hilbert space $\mathcal{H}$ we choose a particular basis.  Any density matrix diagonal in this basis 
belongs to the set of the incoherent states and a given measure of coherence estimates the distance to the closest incoherent
state.  A state which is incoherent in one particular basis $b_{1}$ may become coherent when it is transformed to a different basis
$b_{2}$ via a unitary transformation.  So in the new basis $b_{2}$ this state need not belong to the set of incoherent state 
and hence it cannot be used to measure the coherence. Alternatively a state in the basis $b_{1}$ which has a finite amount of 
coherence will become diagonal and hence incoherent in the $b_{2}$ basis. The set of incoherent states thus completely
depend upon the choice of the basis used and so the coherence is a basis dependent quantity.  Throughout our study on the 
XYZ model discussed in this work we have computed the coherence in the $z$-basis. 
In this section we investigate whether there is any change in the behavior if the coherence in the $x$ and $y$ bases are measured, and contrast them with the results obtained through the computational basis. 

To compute the coherence in the $x$ or the $y$ basis, the thermal density matrix $\rho$ is rotated into its respective basis. After this, the incoherent state $\rho_{d}$ and the state $(\rho + \rho_{d})/2$ corresponding to the rotated density 
matrix are found, and the quantum Jensen-Shannon divergence is calculated in the same way. The entropy $S(\rho)$ is invariant under basis transformation, but on changing the 
basis the incoherent density matrix $\rho_d$  changes and hence $S(\rho_d)$ and $S((\rho +\rho_d)/2)$ depend on the basis.

\begin{figure*}[t]
\begin{center}
\includegraphics[width=150mm]{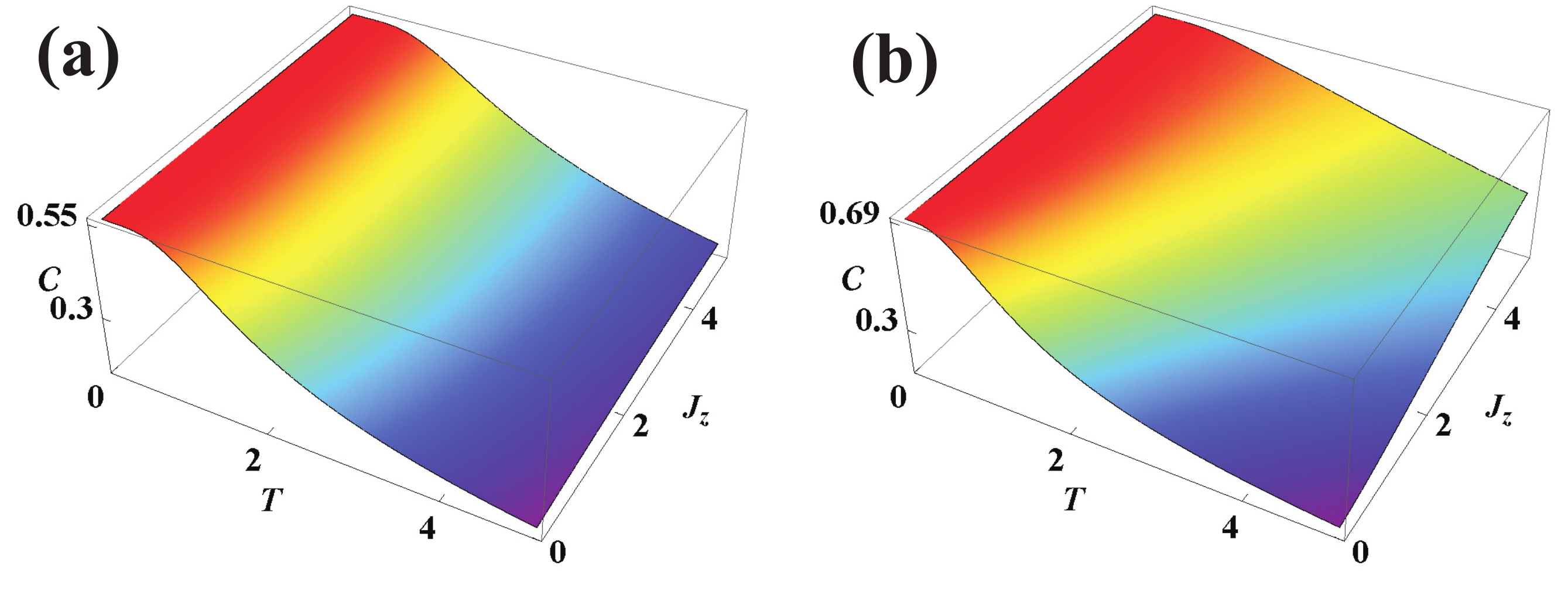}
\end{center}
\caption{Coherence versus temperature and $J_{z}$ for the XYZ model with $D_{z}$ interaction in (a) the $z$-basis and (b) the $x$-basis 
for $J_{x} = -1$, $J_{y} = -0.5$ and $D_{z} = 1.0$.}
\label{fig6}
\end{figure*}

\begin{figure*}
\begin{center}
\includegraphics[width=150mm]{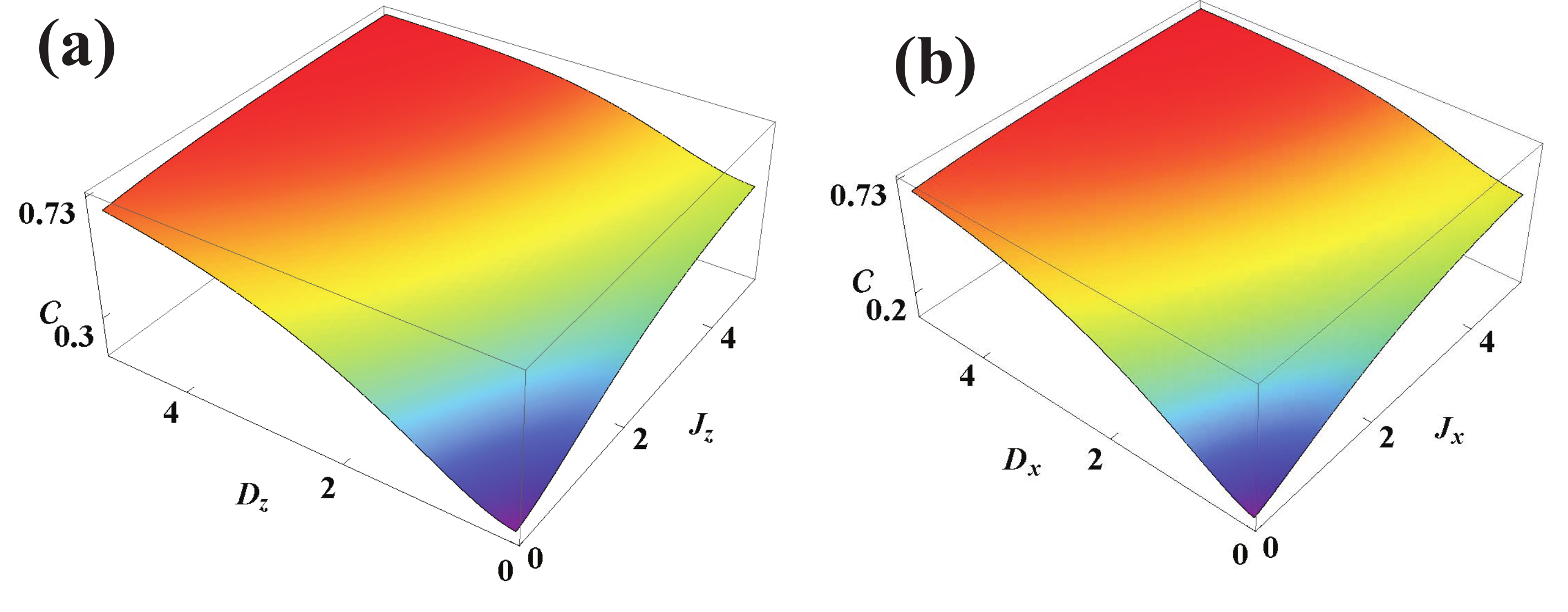}
\end{center}
\caption{(a) Coherence versus $D_z$ and $J_{z}$ in the $x$-basis for $XYZ$ model with $D_{z}$ interaction for $J_{x} = -1.0$, 
$J_{y} = -0.5$ and $T=2.5$. (b) Coherence versus $D_{x}$ and $J_{x}$ in the $z$-basis for the XYZ model with $D_{x}$ interaction
when $J_{y} = -0.5$, $J_{z} =  -1.0$ and $T=2.5$. }
\label{fig7}
\end{figure*}

The coherence in the $x$ and $y$ basis are shown in Figs. \ref{fig6} and \ref{fig7}.  
The thermal density matrix is an X-state (i.e. of the form of (\ref{thermaldmxyzdz}) but with arbitrary diagonal elements)
if the orientation of the DM interaction coincides with the direction of the measurement 
basis.  When the direction of the measurement basis and the DM interaction are different, in general all elements of the density matrix are non-zero. It is well known that the principle of superposition underlies the concept of coherence and a system with more superposition has 
more coherence. Therefore when the direction of the measurement basis and the orientation of the DM coupling are different this gives rise to more coherence in the system in comparison to an X-state where only the diagonal and the anti-diagonal elements are present. 

In Fig. \ref{fig6} we compare the coherence measured in the $z$ and $x$-basis for a XYZ-$D_{z}$ model. 
From the plots we notice that
for a fixed $J_{z}$, quantum coherence decreases with temperature, irrespective of the measurement basis.  But when $J_{z}$ 
is higher, the rate of fall of quantum coherence with temperature is lower when the measurement basis is orthogonal to the direction 
of the DM interaction.  This is because the coherence which is distributed between all the elements can be classified into two classes.  One class contains the diagonal and the antidiagonal elements and the other class contains the remaining terms.  These classes have a different dependence on $J_z$ and temperature which causes this behavior.  When the measurement basis and orientation of the 
DM coupling are the same the density matrix contains only the diagonal and the antidiagonal terms (i.e. an X-state) and thus have
the same $J_{z}$ and temperature dependence and the fall is more uniform.

A comparison of the quantum coherence of the XYZ model as a function of $D_z$ and $J_z$ in the $x$-basis and as a function of $D_x$ and $J_x$ in the $z$-basis is given in Fig. \ref{fig7}.  We find that the dependence of the coherence in Fig. \ref{fig7}(a) and Fig. \ref{fig7}(b)  have a qualitative similarity at a given temperature. We have also calculated the coherence in the $y$-basis for a XYZ-$D_{z}$ model, showing a qualitatively similar behaviour to Fig. \ref{fig7}. 
Thus we observe that the qualitative behaviour of quantum coherence depends on the direction of the 
DM interaction and its relative orientation to the spin-spin coupling terms which are varied. 
 Finally, from our analysis it is clear that whatever basis is used, the coherence in these spin models arise entirely due to the 
correlations between the qubits \cite{radhakrishnan2016distribution} and the value always interpolates between that of the completely ferromagnetic state and the antiferromagnetic state.

%

%
%
\section*{Discussion}
\label{conclusion}

The coherence of the XYZ Heisenberg model with Dzyaloshinsky-Moriya interactions was investigated at both zero temperature and non-zero temperatures.  Two main analysis were performed, one at zero temperature but for an arbitrary number of spins, where an exact solution is possible in limiting cases.  The non-zero temperature case was examined for a minimal two-spin system, under various limits and directions of the DM interaction.  For the zero temperature case, a clear picture of the role of coherence emerged, with the DM antisymmetric exchange and $ J_{x,y} $ symmetric exchange promoting coherence in the system, while the $ J_z $ interaction suppressing it.  The physical picture of this is simple, as the DM interaction in any direction always contains off-diagonal terms in Pauli operators, which tend to create coherence, which is likewise true for the  $ J_{x,y} $ interactions.  The $ J_z $ term favors states which are eigenstates in the $ z $-basis, which naturally do not have coherence.  This can be seen to be true for any number of spins, and the coherence smoothly interpolates between these two limits.  When including temperature effects, we find that quantum coherence decreases with temperature. This can be attributed to a decoherence effect where any superposition instead becomes a mixture.  These are all expected effects and shows that the quantum Jensen-Shannon divergence intuitively captures the nature of coherence, even when extended to a interacting many-body system as the one considered here.

In general, when interaction parameters associated with coherence are increased, the quantum coherence increases.  However, the qualitative behavior depends on the type of the interaction parameter.  When the spin-spin interaction which is aligned along the direction of the DM interaction is varied, the 
saturation value of coherence depends strongly upon the temperature. If the spin-spin interaction being varied is orthogonal to the direction of the DM coupling, then the coherence saturates to almost the same value.  Finally, when the 
anti-symmetric DM coupling strength is changed the coherence tends to change more rapidly than with the symmetric version.  This indicates that the DM interaction influences 
the quantum coherence in a manner which is stronger than the spin-spin coupling.  Another observation is that the temperature at which the coherence falls from its 
maximum value tends to increases with the interaction parameter.  This agrees with the observations for entanglement, but is the opposite to that found for quantum discord.  The reason for this is that the entanglement accounts only for the non-local quantum correlations
of the system, whereas quantum discord takes into consideration the total amount of quantum correlations in the system.  The local 
quantum correlations which are not accounted in the entanglement studies may fall much faster than the rise in non-local quantum 
correlations.  We note that in all the spin models considered in this paper, the local coherence as defined in Ref. \cite{radhakrishnan2016distribution} was zero and the entire 
coherence was due to the interaction between the spins.  Thus the non-local nature of coherence gives rise to a behavior which is 
similar to the one observed for the quantum entanglement.  We also examined the quantum coherence of the models in different bases and found that there is a quantitative and qualitative change in the coherence depending on the choice of the basis. 
The coherence is relatively maximum when the direction of DM interaction is orthogonal to the direction of the measurement basis.  

In addition to revealing some of the coherence properties of a many-body system, our work shows that the quantum Jensen-Shannon divergence is a useful quantity in terms of quantifying the coherence in quantum systems.  As discussed in the introduction one of the main advantages of the coherence measure is that it is formally a distance measure, and thus can be used to distinguish the distribution of coherence in multipartite systems.  It can be potentially applied to a variety of different systems, from naturally occuring many-body systems to synthetic metamaterials \cite{salathe2015digital,johnson2011quantum}.  
Further research in this direction will involve the study of distribution and shareability of coherence in spin models. Earlier studies on a finite sized XXZ model \cite{radhakrishnan2016distribution} have indicated that the 
anisotropy acts as a switch changing the nature
of the system from polygamy to monogamy.  A comparison of this work with these earlier studies will help us to understand the 
role played by DM coupling in the shareability of coherence.   Some other future possibilities include 
the study of spin models with staggered DM interactions \cite{miyahara2007uniform} as well as spin chain models with random exchange interactions \cite{fisher1994random} to understand
the role of coherence in condensed matter systems.  

\vspace{0.5cm}

\begin{acknowledgements}
This work is supported by the Shanghai Research Challenge Fund; New York University Global Seed Grants for Collaborative Research; National Natural Science Foundation of China (Grant No. 61571301); the Thousand Talents Program for Distinguished Young Scholars (Grant No. D1210036A); and the NSFC Research Fund for International Young Scientists (Grant No. 11650110425); NYU-ECNU Institute of Physics at NYU Shanghai; and the Science and Technology Commission of Shanghai Municipality (Grant No. 17ZR1443600).  JS would like to acknowledge the receipt of the Research and Advanced Training Fellowship-2016 by 
The World Academy of Sciences (TWAS) and the NYU-ECNU Institute of Physics at NYU Shanghai for visiting New York University Shanghai.
\end{acknowledgements}

\vspace{0.2cm}
\noindent{\bf Author Contributions} \\ 
RC, JS and TB conceived the problem.  RC did the calculations and  the plots were generated by PM.  
The paper was written by RC, JS and TB.

\vspace{0.2cm}
\noindent{\bf Additional Information} \\
{\bf Competing financial interests:} The authors declare no competing financial interests.  

%


\end{document}